\documentstyle[aps,epsf,floats,preprint,epsfig]{revtex}

\newcommand{\be}{\begin{equation}}
\newcommand{\ee}{\end{equation}}
\newcommand{\beqn}{\begin{eqnarray}}
\newcommand{\eeqn}{\end{eqnarray}}
\newcommand{\bearr}{\begin{array}}
\newcommand{\enarr}{\end{array}}

\begin{document}

\tightenlines \draft

\title{Failure time in the fiber-bundle model with thermal noise and disorder}
\author{Antonio Politi$^{1}$, Sergio Ciliberto$^{2}$ and Riccardo Scorretti$^3$}
\address {$^{1}$ Istituto Nazionale di Ottica Applicata \\
              Largo E. Fermi 6, 50125 Firenze Italy\\
       $^{2}$ Ecole Normale Superi\'eure de Lyon, Laboratoire de Physique\\
           CNRS UMR5672, 46 All\'ee d'Italie, 69364 Lyon, France\\
          $^{3}$ Ecole Centrale de Lyon, 69131 Ecully, France}

\date{\today}
%\twocolumn[
\maketitle
%\widetext
%\twocolumn[
%\widetext
\begin{abstract}
\begin{center}
\parbox{14cm}{
The average time for the onset of macroscopic fractures is analytically and
numerically investigated in the fiber-bundle model with quenched disorder
and thermal noise under a constant load. We find an implicit exact expression
for the failure time in the low-temperature limit that is accurately confirmed
by direct simulations. The effect of the disorder is to lower the energy
barrier.}
\end{center}
\end{abstract}

%]
\narrowtext
\newpage
\section{Introduction}
The onset of fractures in heterogeneous materials has been for a
long time the subject of many studies in the engineering community
for its obvious technological implications. More recently, the
problem has attracted the interest also of the physicist
community, because of its non trivial statistical character.
Fractures are, indeed, genuine transient phenomena for which it is
highly desirable to identify universal laws, but this ambition
contrasts with the lack of general tools capable of dealing with non
equilibrium phenomena. As a consequence, many simplified models
have been proposed in the attempt to capture the relevant
dynamical properties, without pretending to accurately
reproduce the microscopic details.

Theoretical and experimental investigations of fractures are
actually devoted to clarifying many different questions, such as,
e.g., the velocity of propagation, the roughness, and the onset of
precursors. In this paper we are interested in determining the
failure time of a given sample subjected to a constant stress.
This the so-called creep-test, widely used by engineers in order
to estimate the lifetime $\tau$ of a given material as a function
of the applied stress. It would be obviously very desirable to
construct a theory able to predict the failure time upon the
knowledge of a few ingredients and without having to perform
experimental tests under different stress conditions.

Several authors \cite{pomeau,golubovic,Zhurkov,sethna} conjectured
that the fracture is a thermal activated process whose effective
temperature $T_{eff}$ should coincide with the thermodynamic
temperature $T$. Several experimental observations
\cite{pauchard,bonn,noi} seem to indicate that the activation
model proposed by Pomeau predicts correctly the dependence of
$\tau$ on the applied stress. Conversely, all the
experiments\cite{pauchard,bonn,noi} indicate that the effective
temperature in strongly heterogeneous materials can be several
orders of magnitude larger than $T$, or, equivalently, the energy
barrier is smaller than what theoretically predicted.

The need to clarify this problem has led Guarino et al.
\cite{tesi-Guarino,noi-condmat} to suitably modify the
fiber-bundle model, initially introduced \cite{pierce,daniels} as
a purely deterministic model to describe the behaviour of an
ensemble of fibers, all of them subjected to the same load but with
different breaking thresholds \cite{hermann,hemmer,andersen}. In
the original model, upon increasing from zero the applied stress
nothing happens until the weakest fiber breaks. As a consequence
of the first failure, the average stress increases and this may
induce further failures. In practice, it is only when the applied
load is large enough that an avalanche process sets in, giving rise
to a complete failure of the system.

In order to take into account thermal fluctuations, in
Ref.~\cite{noi-condmat,ciliberto,scorretti}, it was argued that
each single fiber can break at any time with a probability per
unit time proportional to the probability of a thermal fluctuation
above the critical length of the given fiber. As a consequence of
thermal fluctuations, the bundle can break for any imposed stress,
exactly as expected in real systems.

The modified fiber-bundle model has been then studied both in the
homogeneous (same breaking threshold for all fibers) and
heterogeneous case, finding that in the former case, the effective
temperature coincides with the thermodynamic temperature
\cite{roux,ciliberto,scorretti}. In heterogeneous systems, it has
been found that disorder contributes to modifying $T$, but it has
not yet been developed a sufficiently general treatment and the
results existing so far do partly conflict with each other and this
prevents drawing definite conclusions.

It is precisely the goal of this paper to develop a general
approach for dealing with heterogeneous systems in the low
temperature limit. We shall show that disorder contributes as a
multiplicative correction that can equivalently be interpreted
either as an amplification of the temperature or a lowering of the
energy barrier. Similar conclusions on the  role of disorder in
the crack activation processes have been reached by other
authors\cite{Arndt}.

In the next section, we briefly recall the results obtained in the
two previous papers that have dealt with the same model. In
section III we derive and solve the dynamical equations that allow
us to determine the scaling behavior for the average failure
time. The last section is devoted to conclusions and an outline of
future perspectives.

\section{Previous results}

Initially, the interest has been devoted to studying the behaviour of
homogeneous bundles, composed of $N$ fibers. In both
Refs.~\cite{roux,ciliberto}, it has been found that, in the limit
of $N\rightarrow \infty$, the average failure time $\tau$ is,
\begin{equation}
\tau = \frac{\sqrt{2\pi T}}{\gamma f_0} \exp 
\left[\frac{(1-f_0)^2}{2T}\right] ,
\label{homo}
\end{equation}
where $T$ is the temperature scaled to the bond energy at the breaking
threshold ($T = \frac{k_B \overline T}{Y\ell}$, where $Y$ is the elastic 
constant, $k_B$ the Boltzmann constant, $\overline T$ the absolute 
temperature and $\ell$ the critical length), $f_0$ is the imposed average 
force (scaled to $Y\ell$), and $1/\gamma$ is the 
time scale of the thermal fluctuations. Thus, in this specific case,
the energy barrier to be overcome in order to break the fiber
bundle is $U = Y\ell^2(1-f_0)^2/2$. Moreover, Roux \cite{roux} showed that
the average failure time of the first fiber is
\begin{equation}
  \tau_1 =  \sqrt{\frac{2\pi}{T}}\frac{1-f_0}{\gamma N}
 \exp \left[\frac{(1-f_0)^2}{2T}\right] .
\label{homo2}
\end{equation}
Accordingly, we see that the exponential factor is the same in
both the expression for $\tau$ and $\tau_1$, indicating that the
activating energy is the same for both processes.

The disordered case is more easily studied under the assumption of a Gaussian
distribution of the breaking thresholds $f$,
\begin{equation}
  P(f) =  \frac{1}{\sqrt{2\pi T_d}} \exp \left[-\frac{(f-1)^2}{2T_d}\right]
\label{diso}
\end{equation}
where the variance $T_d$ measures the amount of quenched
disorder present in the bundle. In order to be precise, one should restrict the
definition of $P(f)$ to positive values, but we shall see in the next section
that in the regime we are interested in, this initial anomaly disappears
immediately without causing any trouble.

With the above assumption, Roux determined again the average failure time of
the first fiber, finding
\begin{equation}
  \tau_1 =  \sqrt{\frac{2\pi}{T+T_d}}\frac{1-f_0}{N}
 \exp \left[\frac{(1-f_0)^2}{2(T+T_d)}\right] .
\label{diso2}
\end{equation}
Accordingly, he concluded that the effect of disorder is to introduce an
additive shift on the effective temperature.

On the other hand, Ciliberto et al.~\cite{ciliberto,scorretti},
performing an analytic approximate calculation, found a
multiplicative correction, namely
\begin{equation}
  \tau =  \tau_0 \exp \left[\frac{(1-f_0)^2}{2T_{eff}}\right] ,
\label{diso3}
\end{equation}
with
\begin{equation}
  T_{eff} = \frac{T}{\left(1 - \frac{\sqrt{2\pi T_d}}{2(1-f_0)} \right)^2}
\label{diso4}
\end{equation}
Since, one cannot control the accuracy of the approximations involved in
the determination of the above formula, it is not possible to discuss a priori
its validity in the small temperature limit, when the analogy with
activation processes becomes more transparent.

\section{Model solution}

Let us start by denoting with $f_a(t)$ the force exerted at time $t$ on a
fiber whose critical force is $f$. According to the original formulation
of the problem, in the presence of thermal noise, the force applied on each
fiber exhibits Gaussian fluctuations around an average value $f_a$.
Therefore, the probability per unit time to break a fiber characterized by
a threshold $f$ is proportional to the probability for a fluctuation
to overcome the assigned threshold, i.e.
\begin{equation}
  G(f-f_a) = \frac{\gamma}{2} \left\{1- \hbox{erf}
  \left[-\frac{(f-f_a)^2}{2T}\right] \right\}
\label{gaus0}
\end{equation}
where $T$ is the working temperature, while $\gamma$ is a constant fixing
the time scale for the process.
In the small temperature limit, we will see that the most relevant
contribution to the fiber breakdown occurs in the tail of the distribution,
where we can approximate the error function with a Gaussian. Accordingly, we
assume that
\begin{equation}
  G(f-f_a) = \frac{\gamma}{f-f_a}\sqrt{\frac{T}{2\pi}}
\exp \left[-\frac{(f-f_a)^2}{2T}\right] .
\label{gaus1}
\end{equation}
Let us now introduce the relevant dynamical variable, i.e. the distribution
$Q(f,t)$ of unbroken bonds at time $t$ ($Q(f,0) = P(f)$).
The fraction of broken bonds is, therefore,
\begin{equation}
  \Phi(t) \equiv 1 - \int_{-\infty}^{+\infty} df Q(f,t) ,
\end{equation}
and the average force $f_a$ exerted on each fiber at time $t$ is
\begin{equation}
  f_a = \frac{f_0}{1-\Phi}
\label{fa}
\end{equation}
where $f_0$ is the initial average force. The definition of the model is
completed by the dynamical equation for $Q(f,t)$
\begin{equation}
  \dot Q(f,t) = -Q(f,t)G(f-f_a)  .
\label{tot}
\end{equation}
A similar model has been studied in Ref.~\cite{NTG95} in connection to the
investigation of seismic activation, the main difference being that in their
case, the breaking rate is given rather than being self-consistently 
determined. It is precisely the resulting time dependence of $f_a$ (determined 
by the integral of $Q$ over all $f$ values) which makes Eq.~(\ref{tot}) 
difficult to solve.

\begin{figure}[tcb]
\begin{center}
\includegraphics*[width=9cm]{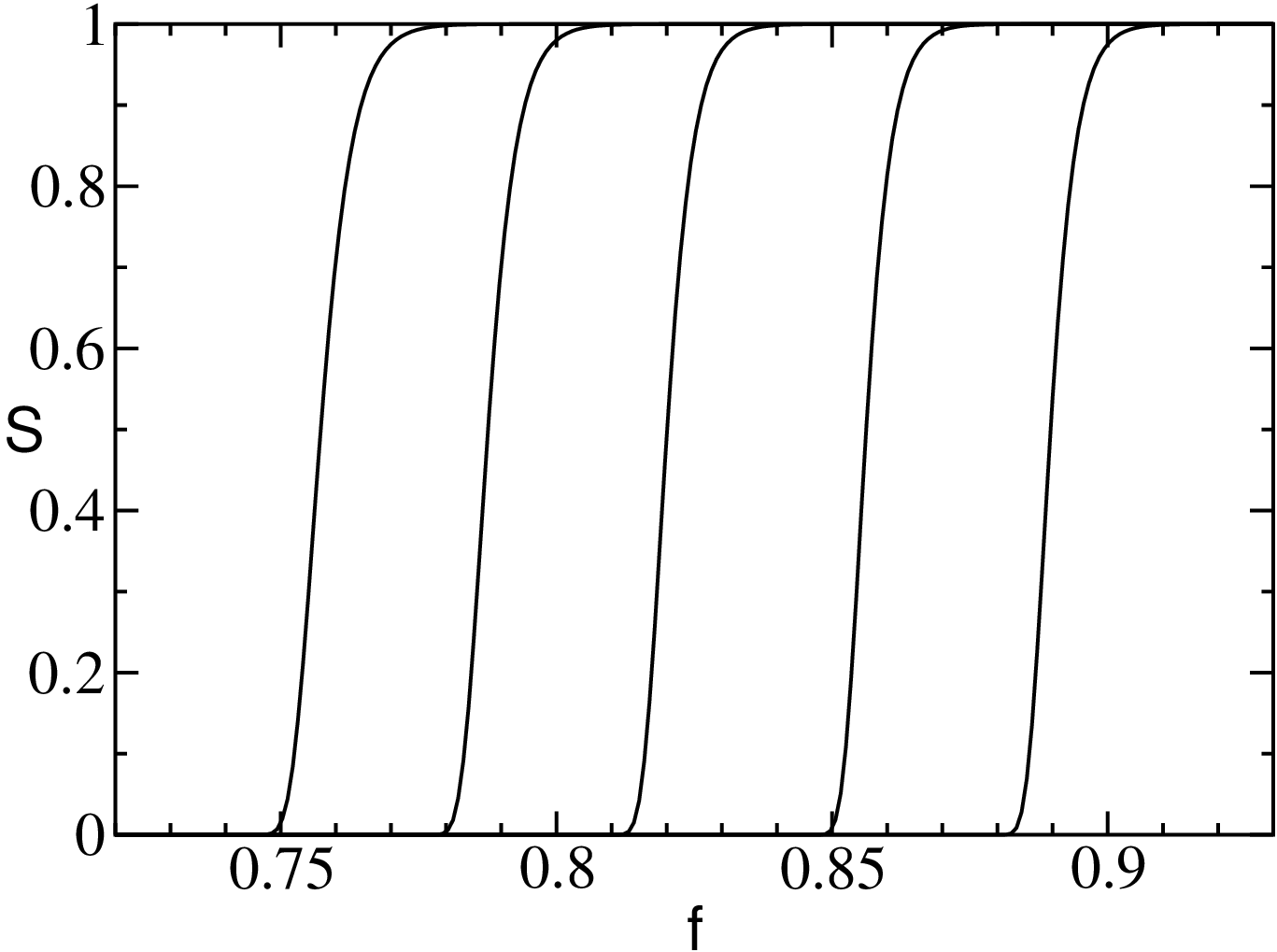}
\caption{Position of the front $S$ at times $6.6\, 10^{13}$, $6.6\, 10^{16}$,
 $5.5\, 10^{19}$, $6.6 \, 10^{13}$, $8.5\, 10^{21}$, $3.3 \, 10^{23}$ (from
left to right) in a simulation with $T= 10^{-3}$, $T_d =10^{-2}$ and
$\gamma=1$. All the variables reported in this and in the following figures
are dimensionless.}
\label{fronts}
\end{center}
\end{figure}

Before passing to the analytical calculations, let us discuss the numerical
integration of Eq.~(\ref{tot}). We find it convenient
to introduce the variable
\be
 S(f,t) =  \frac{Q(f,t)}{P(f)}
\ee
representing the fraction of unbroken bonds at time $t$ per class of fibers
with thresholds between $f$ and $f+df$. In fact, the evolution of $S$
provides an insightful representation of the fracture process.
As one might have expected, we see in Fig.~\ref{fronts} that the breakdown
of the bundle starts from the weaker fibers to progressively affect the
more robust ones. Less obvious, is that the fracture appears to proceed as
a moving front with constant shape. Moreover, the temporal spacing of the
various fronts reported in Fig.~\ref{fronts} reveals a progressive slowing
down of the evolution. This latter feature will turn out to be the crucial
point for understanding the scaling properties of the whole process.

\begin{figure}[tcb]
\begin{center}
\includegraphics*[width=9cm]{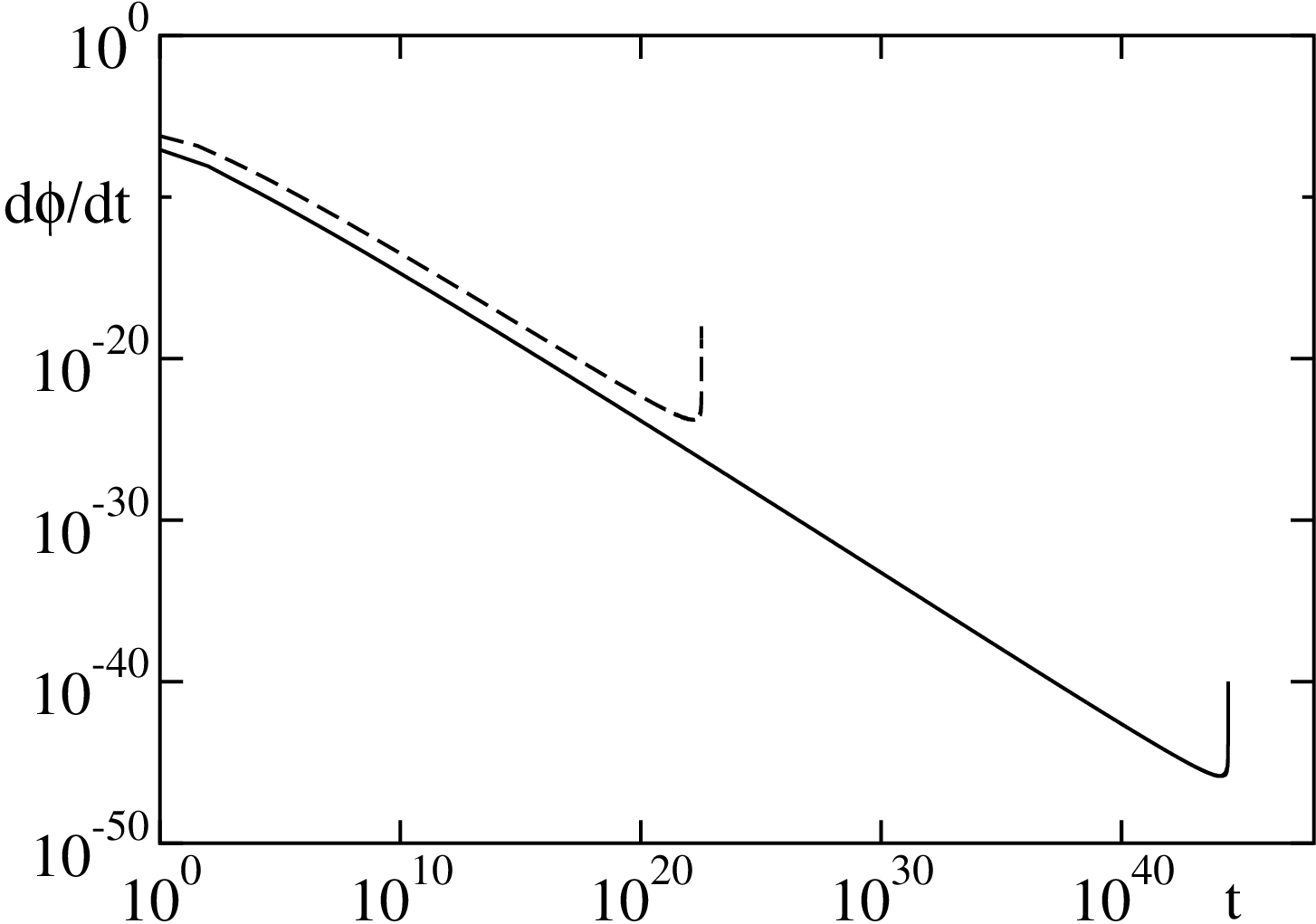}
\caption{Time derivative of $\Phi$ versus time for $T_d = 10^{-2}$ and
two different values of the temperature, $T= 5. 10^{-4}$ (solid curve),
$T = 10^{-3}$ (dashed curve).}
\label{deriv}
\end{center}
\end{figure}

The increasing slowness of the bond breakdown is better revealed
by looking  at the time derivative of $\Phi$. The monotonous
decrease of $\dot \Phi$ preceding the final macroscopic fracture
(see Fig.~\ref{deriv}) indicates that one cannot estimate the
average breaking time $\tau$ by limiting oneself to follow the
initial stages of the process.

A yet clearer description of the breakdown process is obtained by
formally interpreting $S(f)$ as the integral of some probability
distribution $R'(f)$ (i.e. $dS/df = R'(f)$). This allows also a
straightforward identification of the step region, where the
ongoing breakdowns are concentrated at a given time. More
interesting, we find that the shape of $R'(f)$ is independent of
$T$ in the slowest evolution region (i.e., where most of the time is
spent before the final breakdown).
\begin{figure}[tcb]
\begin{center}
\includegraphics*[width=9cm]{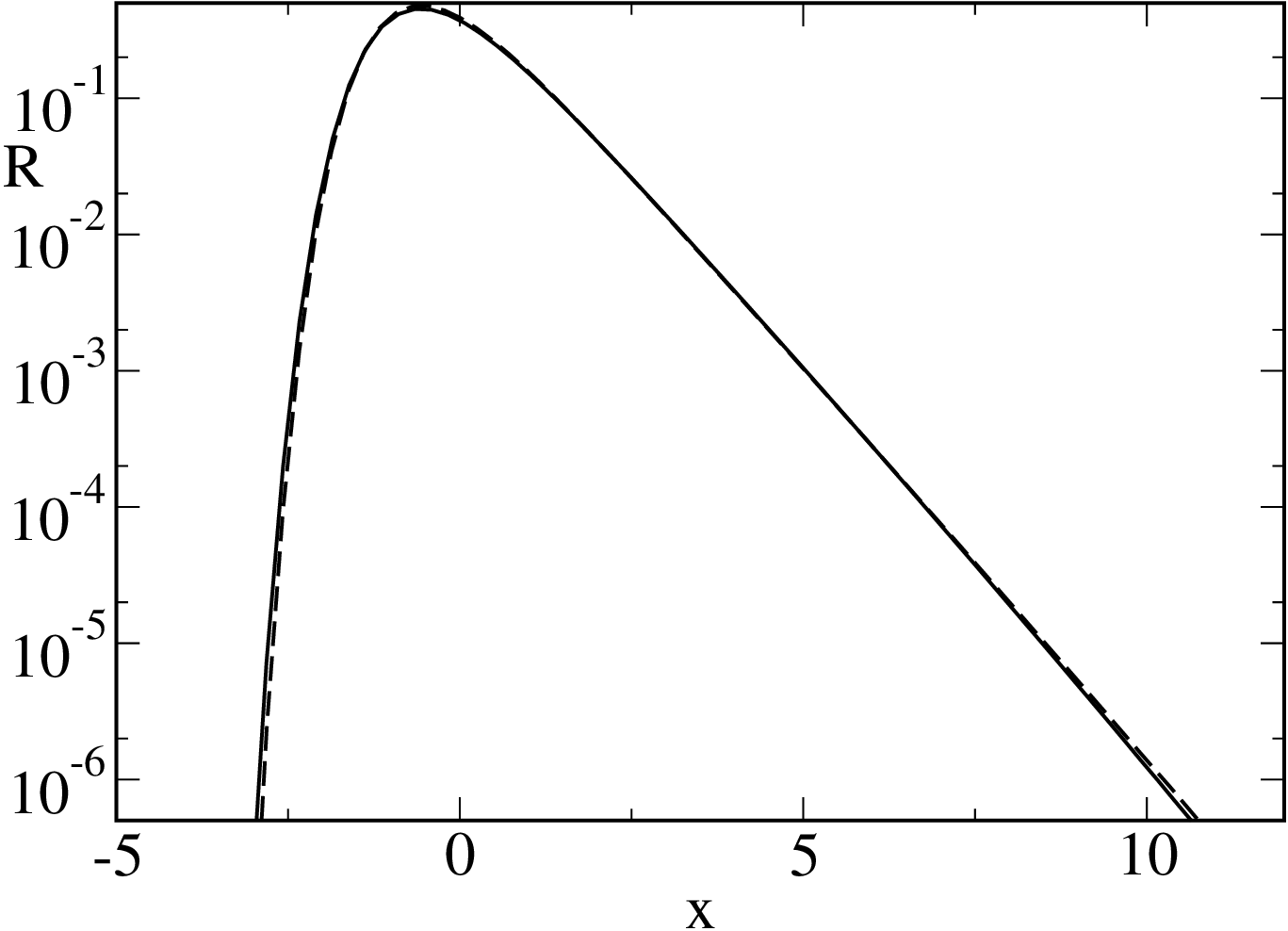}
\caption{The ``probability'' $R(x)$ scaled to unit variance and shifted
around the center of mass for the same values as in the previous figure and
the same notations (see the body of the text for the definition of $x$).}
\label{over}
\end{center}
\end{figure}
This can be appreciated in Fig.~\ref{over}, where we have plotted 
$R(x) = \sigma_r R'(f)$ for two different temperature values, after shifting 
the distribution around the average value $\overline f$ and scaling $f$ to 
the r.m.s. $\sigma_r$ (i.e.  $x = (f-\overline f)/\sigma_r$).

Besides observing the independence of the shape on the temperature, notice
also the strong similarity with distributions obtained for
extreme-value statistics \cite{Holdsworth,gamble,bouchaud}. This is
certainly not a surprise, since the tail of $R(x)$ consists of events
that, over time, proved to be anomalous. The accuracy of the data allows
us to show that in this case the Gumbel's distribution \cite{gamble,bouchaud}
fits precisely the $R(f)$. This specific shape of the $R'(f)$ and its  scale
invariance deserve further investigations, but here we are more
interested in describing the temporal evolution of the fracture
process. To this goal, it is more important to notice that the
standard deviation $\sigma_r$ of $R'(f)$ goes to 0 linearly with $T$.
\begin{figure}[tcb]
\begin{center}
\includegraphics*[width=9cm]{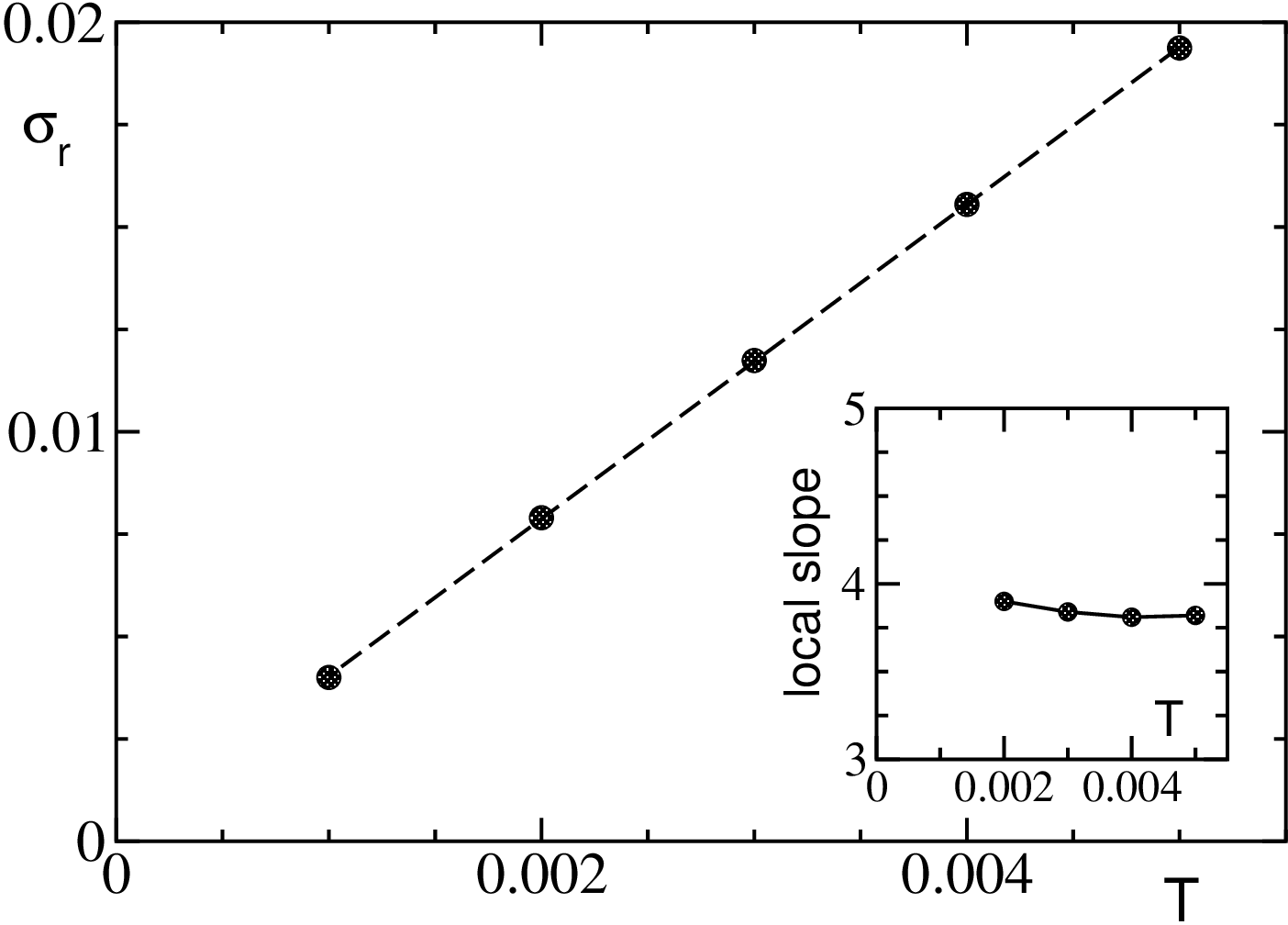}
\caption{The standard deviation $\sigma_r$ of $R'(f)$, computed when the time
derivative of $\Phi$ is minimum, versus the temperature $T$ for fixed
disorder $T_d = 10^{-2}$. In the inset, we can appreciate the small
deviations from a purely linear behaviour.}
\label{large}
\end{center}
\end{figure}
This can be clearly seen in Fig.~\ref{large}, where we have reported
$\sigma_r$ versus $T$ for $T_d = 10^{-2}$ (there, it can also be seen that
the proportionality constant is approximately equal to 4). Notice that the
linear dependence on $T$ is quite a fast decrease, as thermal fluctuations
are on the order of $\sqrt{T}$.
This suggests that a good (asymptotically exact for $T \to 0$)
approximation consists in assuming a Heaviside shape for $S(f)$.
Such an approximation has also the advantage of parametrizing an a
priori infinite-dimensional object such as $S(f)$ with a single
variable: the position of the step $f_s$. Equipped with such an
assumption, $Q(f,t)$ can be approximated with a Gaussian truncated
below some threshold $f=f_s$. Notice that this differs from the
hypothesis formulated in Ref.\cite{ciliberto,scorretti}, where it
was assumed that $Q(f,t)$ remains unchanged for $f>1$ while it
decreases linearly to 0 for $f<1$ with a slope to be determined
self-consistently.

By integrating
Eq.~(\ref{tot}) over $f$, we obtain the one-dimensional differential equation
\begin{equation}
 \dot \Phi = \frac{\gamma}{2\pi}\sqrt{\frac{T}{T_d}} \int_{f_s}^\infty
                 \frac{df}{f-f_a}
                 \exp \left[-\frac{(1-f)^2}{2T_d} \right]
                 \exp \left[-\frac{(f-f_a)^2}{2T} \right]
\label{single}
\end{equation}
where the dependence on $\Phi$ in the r.h.s. is contained in $f_a$
(see Eq.~(\ref{fa})) and in $f_s$ through the following obvious equation
\begin{equation}
 \Phi =  \frac{1}{\sqrt{2\pi T_d}}\int_{-\infty}^{f_s} df
        \exp \left[-\frac{(1-f)^2}{2T_d} \right]
\label{nor}
\end{equation}
Upon suitably rewriting the product of two Gaussians in Eq.~(\ref{single}),
we obtain
\begin{equation}
 \dot \Phi = \frac{\gamma}{2\pi}\sqrt{\frac{T}{T_d}}
       \exp \left[-\frac{(1-f_a)^2}{2(T+T_d)} \right]
       \int_{f_s}^\infty \frac{df}{f-f_a}
 \exp \left[-\frac{(f-f_b)^2}{2T_b} \right]
\label{sinri}
\end{equation}
where
\begin{equation}
 f_b = \frac{T+f_aT_d}{T+T_d}
\label{fb}
\end{equation}
and
\begin{equation}
 T_b = \frac{T T_d}{T+T_d} .
\label{tb}
\end{equation}
Several observations are now in order. The dependence on the
temperature is very different in the two exponentials. The
variance in the term out of the integral is the sum of the true-
and of the disorder-temperature. This is the contribution that was
already singled out by Roux in Ref.~\cite{roux}. The second term,
instead, exhibits a dependence as if the two temperatures were in
parallel. Now, it is important to establish which term is the
leading one in determining the relevant time scale. As long as
$f_b > f_s$, the exponential integral is of order 1 and the
evolution is controlled by the first term. However, this is not
what happens (at least except for the very first and last stages)
in the limit of very small $T$. To discuss this point, we must
keep in mind all the various $f$'s that are involved in the
process at a generic time $t$, starting from $f_a(t)$, the average
force applied to each unbroken fiber, going to $f_s(t)$, the
threshold of the weaker fiber to break, and to $f_b(t)$ the most
numerous fibers to break (if still alive).

\begin{figure}[tcb]
\begin{center}
\includegraphics*[width=9cm]{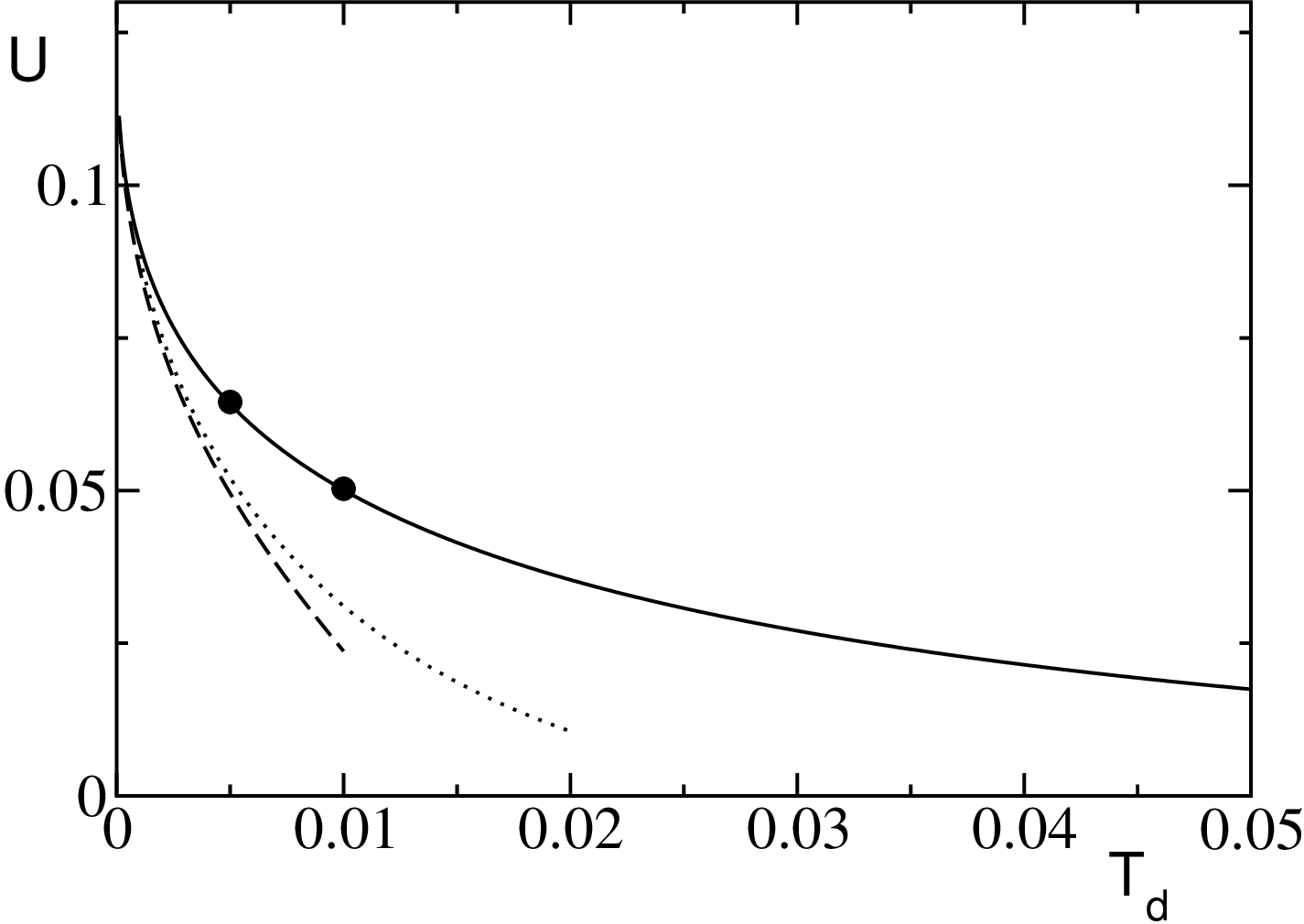}
\caption{The effective barrier energy $U$ as determined from the
numerical solution of the equation (\ref{second}) (solid line).
The two circles refer to the extrapolated value of $U$ from direct
numerical simulations.The dashed line refers to the perturbative
formula (\ref{third}), while the dotted line corresponds to the
approximated solution (\ref{diso3}).} \label{last}
\end{center}
\end{figure}

If $T$ is very small, it is by far easier to break the few fibers
whose threshold is just above the applied force $f_a$ than the
very many fibers with high threshold. This implies that in the
very beginning of the fracture process, it is generated a gap
between the force needed to break the weakest fibers and the
average applied force, leading to a picture analogous to that for
the homogeneous case. Under such conditions, $f_b < f_s$ and the
integral is dominated by the amplitude of the integrand at the
left extremum $f_s$ of the integration domain. Upon computing the
leading contribution to the integral in Eq.~(\ref{sinri}), we can
write
\begin{equation}
 \dot \Phi = \frac{\gamma \sqrt{TT_d}}{2\pi(f_s-f_a)(f_s-f_b)}
       \exp \left[-\frac{(1-f_a)^2}{2(T+T_d)}
 +\frac{(f_s-f_b)^2}{2T_b} \right] .
\label{sinri2}
\end{equation}
An upper bound to $\tau$ can be obtained by determining the maximum of
\begin{equation}
 \tau (\Phi) = 1/\dot \Phi(\Phi) .
\end{equation}
Such an estimate would be exact only in the case of a constant derivative:
although we have seen in Fig.~\ref{deriv} that this is not the case, it is
nevertheless true that most of the time is spent near the minimum of the
derivative, so that we can expect that the above estimate is rather accurate.
With no pretense of estimating prefactors, let us pay attention only at
the exponential factors in the above equation. In the small $T$ limit, the
first contribution is negligible, and thus we write
\begin{equation}
 \ln \tau  \approx  \frac{(f_s^*-f_a^*)^2}{2T}  \equiv \frac{U}{T}
\label{tscale2}
\end{equation}
where $f_s^*$ and $f_a^*$ are the $f_s$- and $f_a$-values yielding
the minimum $\dot \Phi$. $U \equiv (f_s^*-f_a*)^2/2$ can be
interpreted as the effective energy barrier to be overcome in the
activation process to give rise to the final breakdown. It is
instructive to notice that $U$ is smaller than the height in the
homogeneous case ($(1-f_0)^2/2$) for two reasons: (i) $f_0$
increases to $f_a^*$ as a consequence of the initial
``easy'' ruptures that occur on short time scales; (ii) the most
populated class of thresholds ``f=1'' decreases to $f_s*$, the
critical force above which the process starts accelerating giving
eventually rise to an avalanche. The $\Phi^*$ value corresponding
to the maximum of $\tau(\Phi)$ (and, in turn, the values $f_s^*$
and $f_a^*$) can be determined from the zero of the derivative of
$\tau(\Phi)$. From Eq.~(\ref{tscale2}), taking into account
Eq.~(\ref{nor}), one obtains
\begin{equation}
 (1-\Phi^*)^2 = \frac{f_0}{\sqrt{2\pi T_d}} \exp \left[
       \frac{(1-f_s^*)^2}{2T_d}\right] .
\label{second}
\end{equation}
Eq.~(\ref{second}), together with Eq.~(\ref{nor}), determines the
critical value $\Phi^*$ and thus the effective height $U$ of the
energy barrier. In Fig.~\ref{last} we have plotted $U$ versus
$T_d$. As expected, in the limit $T_d \to 0$, $U$ converges to
1/8, the height in the absence of disorder for $f_0=1/2$ (the
value fixed in our numerical simulations). The decrease of $U$
with $T_d$ confirms that the presence of disorder helps the
fracture process, making it more probable. In the limit of small
$T_d$, $\Phi^*$ tends to 0 and one can perform a perturbative
calculation, obtaining
\begin{equation}
  U = \frac{(1-f_0)^2}{2} - (1-f_0)\sqrt{T_d} \left\{
      \left[ 2\ln \left(\frac{f_0}{\sqrt{2\pi T_d}}\right) \right]^{1/2} +
      \left[ 2\ln \left(\frac{f_0}{\sqrt{2\pi T_d}}\right) \right]^{-1/2}
       \right\} .
\label{third}
\end{equation}
The two terms contributing to the deviation from the homogeneous
case arise, respectively, from the decrease of $f_s^*$ below 1 and
the increase of $f_a$ above $f_0^*$. Both corrections are
approximately of the same order, i.e. $\sqrt{T_d}$. It is only by
looking at the logarithmic correction that we can conclude that
the former contribution is the largest one. It is presumably the
presence of such corrections that makes the validity range of this
perturbative calculation so small, as it can be seen by looking at
the dashed line in Fig.~\ref{last}. In the same figure, we have
reported also the analytic solution (\ref{diso3}) obtained in
Ref.~\cite{ciliberto}: its closeness to the perturbative solution
suggests that the result is rather robust against approximations
made on the shape of $Q(f,t)$.

\begin{figure}[tcb]
\begin{center}
\includegraphics*[width=9cm]{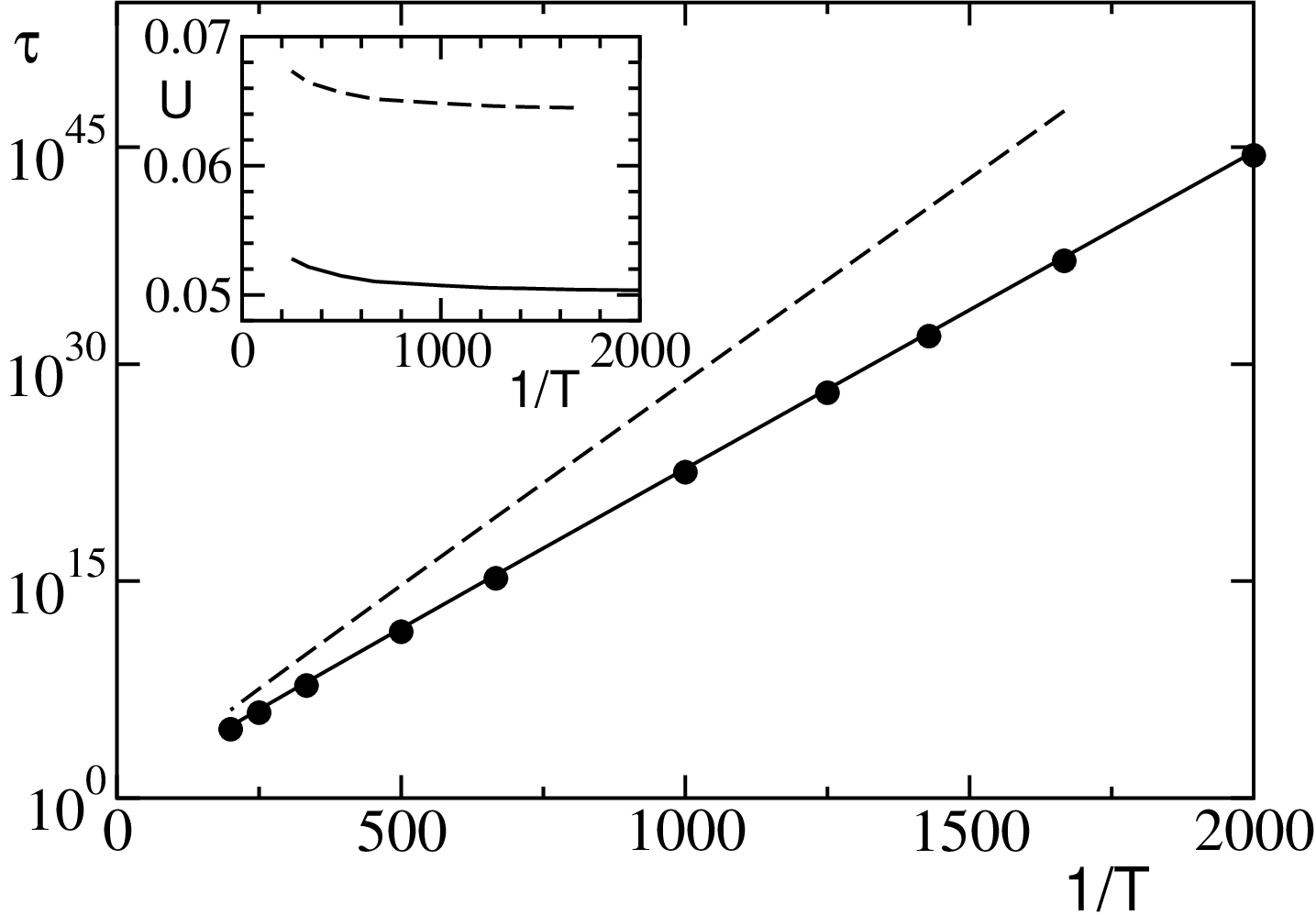}
\caption{The rupture time versus the inverse temperature for two different
values of the disorder temperature as determined from the integration of
the simplified one-dimensional equation (\ref{single}): the solid and dashed
lines refer to $T_d=10^{-2}$, and $T_d=5 10^{-3}$. Circles correspond to
the integration of the full equation. In the inset we report the energy
barrier $U$ determined as the local logarithmic derivative of $\tau$ with
respect to $1/T$}
\label{scal}
\end{center}
\end{figure}

We conclude the analysis, by comparing these theoretical predictions with
the outcome of numerical simulations performed both by integrating the
one-dimensional Eq.~(\ref{single}) and the original model. In Fig.~\ref{scal},
we have plotted the rupture time versus $1/T$ for two different values of
the disorder temperature $T_d$. The rather clean linear behaviour confirms
the scaling behaviour expected for an activation process. In fact, it is
necessary to look at the local logarithmic derivative of $\tau$ (which
corresponds to $U$) to see
deviations from linearity (see the inset) and even this analysis indicates
that deviations from linearity vanish for $T \to 0$.
By comparing the full circles with the solid line, we can instead appreciate
the validity of the truncated-Gaussian approximation, since the circles refer
to the integration of the full model, while the solid line arises from the
one-dimensional approximation.

We are now in the position to compare the value of $U$, extrapolated from
numerical simulations, with the theoretical prediction plotted in
Fig.~\ref{last}. The fact that the two circles fall precisely on top of
the theoretical curve further confirm the validity of the whole approach.

\section{Conclusions and perspectives}

The analytical treatment developed in this paper confirms the claim that
the presence of disorder contributes to increasing the effective temperature
of a sample subject to a constant load. Equivalently, but perhaps more
physically, one can state that disorder renormalizes the barrier height to
be overcome in order to give rise to a macroscopic failure of the fiber
bundle. This scenario can be understood by noticing that the fracture evolves
through a sequence of many irreversible processes. After the failure of the
weakest fibers, the system cannot any longer come back to its initial state,
while, at the same time, the energy barrier has lowered. A correct estimation
of the time scale for observing the onset of a macroscopic failure is
obtained by determining the time scale for the slowest of such intermediate
steps.

 From the way this result has been obtained, there is no reason to suspect that
it follows from some peculiarity of the fiber-bundle model with quenched noise.
As, indeed, suggested by experimental results, it is natural to conjecture
that the presence of noise lowers the energy barrier also in more realistic
set-ups. It becomes desirable now to implement more general tools to go
beyond mean field models.

%\acknowledgements
 One of us (AP) wishes to thank the ENS Lyon for
the invitation that has allowed starting this work.


\begin{thebibliography}{99}


  \bibitem{pomeau}  Y. Pomeau,
%  {\it Brisure spontan\'{e}e des cristaux bidimensionnel courb\'{e}s},
  C. R. Acad. Sci. Paris, vol. 314 II,
  553-556 (1992).

\bibitem{golubovic}  Golubovic L. and Feng S.,
   Phys. Rev. A, 43, 5223 (1991).

\bibitem{Zhurkov} S. N. Zhurkov, Int. J. Fracture, 1, 311, (1965)

\bibitem{sethna} A.Buchel, J.P. Sethna, Phys Rev. E, 55, 7669
(1997).

 \bibitem{pauchard}  L. Pauchard and J. Meunier, Phys. Rev. Lett., 70, 3565, (1993).

\bibitem{bonn}  D. Bonn, H. Kellay , M. Prochnow, K. Ben-Djemiaa and J.
  Meunier, Sciences, 280, 265 (1998).

 \bibitem{noi}  A. Guarino, A. Garcimart\'{i}n and S. Ciliberto, Europhys. Lett., 47, 456
  (1999).

\bibitem{tesi-Guarino} A. Guarino, Phd Thesis, Ecole Normale Sup\'erieure
de Lyon, France

\bibitem{noi-condmat} A. Guarino, R. Scorretti, S. Ciliberto,
cond-mat/9908329

\bibitem{pierce}  F.T. Pierce, J. Textile Industry, (1926).

\bibitem{daniels} H. E. Daniels, Proc. R. Soc. London, Ser. A,
183, 658 (1945).

\bibitem{hermann}  H.J Hermann, S. Roux, {\it Statistical models for the
  fracture of disordered media}, North-Holland, Amsterdam (1990).

 \bibitem{hemmer}  C. Hemmer. A. Hansen,  Journ. of Applied Mech.,
   59, 909 (1992).

  \bibitem{andersen}  J. V. Andersen, D. Sornette, K.T. Leung, Phys.
  Rev. Lett. 78, 2140 (1997).

%\bibitem{Coleman} B.D. Coleman, J. Appl. Phys., 29, 968 (1958).

%\bibitem{Phoenix} S.L.Phoenix, L. J. Tierney, Eng. Fract. Mech., 18, 193 (1983).

\bibitem{ciliberto} S. Ciliberto, A. Guarino, R. Scorretti,
Physica D, 158, 83 (2001).

\bibitem{scorretti}  R. Scorretti,S. Ciliberto, A. Guarino,
Europhys. Lett. 55, 626 (2001).

\bibitem{Arndt} P. Arndt and T. Nattermann, Phys. Rew. B, 73, 134
(2001).

\bibitem{roux} S. Roux, Phys. Rev. E 62, 6164 (2000).

\bibitem{NTG95} W.I. Newman, D.L. Turcotte, and A.M. Gabrielov, Phys. Rev. E
{\bf 52}, 4827 (1995).

\bibitem{Holdsworth} S.T. Bramwell, P. C. W.Holdsworth, J. F. Pinton,
Nature, 396, 552 (1998).

\bibitem{gamble} E. J. Gumbel, {\it Statistics of Extremes}
(Columbia University Press, New York 1958).

\bibitem{bouchaud} J.P. Bouchaud, M. M\'ezard, J. Phys. A 30, 7997
(1997).



\end{thebibliography}
\end{document}